\documentclass[amsmath,amssymb,showkeys, showpacs]{revtex4}
\usepackage{graphicx}
\usepackage{accents}
\usepackage{cancel}
\expandafter\let\csname equation*\endcsname\relax
\expandafter\let\csname endequation*\endcsname\relax
\usepackage{amsmath,amssymb,amsthm,amsfonts,color}

\newcommand{\ba}{\begin{array}}
\newcommand{\ea}{\end{array}}

\numberwithin{equation}{section}

\newcommand{\bea}{\begin{eqnarray}} 
\newcommand{\eea}{\end{eqnarray}}  

\newcommand{\bg}{\begin{gathered}}
\newcommand{\eg}{\end{gathered}}

\newcommand{\vs}{\vskip0.1true in \noindent }
\newcommand*\pFq[2]{{}_{#1}F_{#2}} %Hypergeometric function

\theoremstyle{definition}

\theoremstyle{remark}

\begin{document}

% ---------------------------
%\hspace*{4 in}CUQM-150\\
\hspace*{4 in} (Revised) file name: KBS.tex \today
\vspace*{0.4 in}
% ---------------------------
\title{A Note on the Generalized and Universal Associated Legendre Equations}

\author{Keegan L. A. Kirk, Kyle R. Bryenton and Nasser Saad}
\address{School of Mathematical \& Computational Sciences, 
University of Prince Edward Island, 550 University Avenue,
Charlottetown, PEI, Canada C1A 4P3.}\email{kkirk@upei.ca,kbryenton@upei.ca,nsaad@upei.ca}

\begin{abstract}
\noindent A class of second-order differential equations commonly arising in physics applications are considered, and their explicit hypergeometric solutions are provided. Further, the relationship with the Generalized and Universal Associated Legendre Equations are examined and established. The hypergeometric solutions, presented in this work, will promote future investigations of their mathematical properties and applications to problems in theoretical physics.
\end{abstract}
\maketitle
%%%%%%%%%%%%%%%%%%%%%%%%%%%%%%%%%%%%%%
\section{Introduction}\label{intro}
%%%%%%%%%%%%%%%%%%%%%%%%%%%%%%%%%%%%%%

\noindent Recently, the Universal Associated Legendre Polynomials 
\begin{align}\label{sec1eq1}
P_{\ell'}^{m'}(r)&=\sqrt{\frac{(2\ell'+1)(\ell'-m')!}{2\Gamma(\ell'+m'+1)}}(1-r^2)^{m'/2}\sum_{\nu=0}^{[\frac12(\ell'-m')]}\dfrac{(-1)^\nu\Gamma(2\ell'-2\nu+1)}{2^{\ell'}\nu!(\ell'-m'-2\nu)!\Gamma(\ell'-\nu+1)}r^{\ell'-m'-2\nu},
\end{align}
has been the subject of many interesting studies \cite{F2014, F2015,C2016,F2016,Y2017,D2017,W2017}. These polynomials are  solutions to the differential equation (see \cite{F2015} and the references therein):
\begin{align}\label{sec1eq2}
\left(1-r^2 \right)\frac{d^2P_{\ell'}^{m'}(r)}{dr^2}& -2\,r \frac{dP_{\ell'}^{m'}(r)}{dr} + \left(\lambda- \frac{m^2}{1-r^2}- \frac{a+br+cr^2}{1-r^2} \right)P_{\ell'}^{m'}(r)= 0,\quad (-1\le r\le 1),\quad a,c\in \mathbb R,
\end{align} 
in which $b=0,~m'=\sqrt{a+c+m^2},~\lambda=\ell'(\ell'+1)-c$, $\ell'=m'+n, ~n=0,1,2,\dots$. Through partial-fraction decomposition of the rational coefficient of the $P_{\ell'}^{m'}(r)$ term, equation \eqref{sec1eq2} is shown to be a slight modification of the well-known Generalized Associated Legendre Equation \cite{bateman,k1957a, k1957b,k1958,M1958a,M1958b,k1958a,k1958b,k1958c,k1958d,V1994,Vi:Fe: 2001}
\begin{align}\label{sec1eq3}
\left(1-r^2 \right)\frac{d^2F(r)}{dr^2}& -2\,r \frac{dF(r)}{dr} + \left(k(k+1)- \frac{n^2}{2\left(1+r\right)}- \frac{m^2}{2 \left(1-r\right)} \right)F(r) = 0.
\end{align}
The differential equation \eqref{sec1eq3} was introduced first by H. Bateman in his analysis of the harmonic equations \cite[page 389]{bateman}. Following Bateman's work, this was later intensively studied  in a series of research articles by Kuipers and Meulenbeld for complex-valued parameters $k,m$, and $n$ \cite{k1957a, k1957b,k1958,M1958a,M1958b,k1958a,k1958b,k1958c,k1958d}. The recent book of Virchenko and Fedotova \cite{Vi:Fe: 2001} was devoted to the subsequent development of
the theory of the Generalized Associated Legendre Functions and their applications. 
\vs
As we shall prove in the present work, both differential equations \eqref{sec1eq2} and \eqref{sec1eq3} are members of a more general class of differential equations characterized by
\begin{equation}\label{sec1eq4}
\left( r-\xi_1 \right) \left(\xi_2-r \right)\frac{d^2F(r)}{dr^2} + \left( a_1 r + b_1 \right) \frac{dF(r)}{dr} + \left( \lambda + \frac{a_2 r + b_2}{\left( r-\xi_1 \right) \left(\xi_2-r \right)} + \frac{ a_3 r^2 + b_3 r + c_3 }{\left( r-\xi_1 \right) \left(\xi_2-r \right)} \right)F(r) = 0,
\end{equation}
where  $a_j,b_j,j=1,2,3, ~c_3$ are real parameters and $\xi_1<r<\xi_2$.
\vs
The exact solutions of the differential equation \eqref{sec1eq4} are given, along with their relations to the published solutions of the Generalized and Universal Associated Legendre Differential Equations \eqref{sec1eq2} and \eqref{sec1eq3}. New solvable classes of differential equations useful for the analysis of quantum systems are obtained \cite{sharma,khan,CHSD:2010}.
%%%%%%%%%%%%%%%%%%%%%%%%%%%
\section{Exact solutions}\label{sec2}
%%%%%%%%%%%%%%%%%%%%%%%%%%%
\noindent The differential equation \eqref{sec1eq4} has three regular singular points, $r \in \{ \xi_1 , \, \xi_2 , \, \infty \}$ with exponents $\mu_1$, $\mu_2$, and $\mu_\infty$, respectively, determined as the roots of the indicial (quadratic) equations:
\begin{align}
\mu_1^2&+\left(\dfrac{a_1\xi_1+b_1}{\xi_2-\xi_1}-1\right)\mu_1+\dfrac{a_3 \xi _1^2+(a_2+b_3) \xi _1+b_2+c_3}{(\xi_2-\xi_1)^2}=0,\label{sec2eq1}\\
\mu_2^2&-\left(\dfrac{a_1\xi_2+b_1}{\xi_2-\xi_1}+1\right)\mu_2+\dfrac{a_3 \xi _2^2+(a_2+b_3) \xi _2+b_2+c_3}{(\xi_2-\xi_1)^2}=0,\label{sec2eq2}\\
\mu_\infty^2&+(1+a_1)\mu_\infty+a_3-\lambda=0.\label{sec2eq3}
\end{align}
According to the classical theory of ordinary differential equations \cite{earl}, equation \eqref{sec1eq4} is reducible to the hypergeometric  equation. To this end, the  general solutions  of \eqref{sec1eq4} take the form
\begin{equation} \label{sec2eq4}
F(r)=\left( r - \xi_1 \right)^{\mu_1}\left( \xi_2-r \right)^{\mu_2} f\left( r \right),
\end{equation}
where the exponents $\mu_1$ and $\mu_2$ are evaluated using \eqref{sec2eq1} and \eqref{sec2eq2} respectively. The substitution of \eqref{sec2eq4} into equation \eqref{sec1eq4} yields the following hypergeometric-type equation for the function $f\equiv f(r)$:
\begin{align} \label{sec2eq5}
(r - \xi_1)(\xi_2-r)\frac{d^2f }{dr^2}&+\bigg[(a_1 - 2\mu_1 - 2\mu_2) \,r+b_1
+ 2\mu_2\xi_1+2\mu_1\xi_2\bigg]\frac{df }{dr}+\bigg[\lambda -a_3- (\mu_1 + \mu_2)(\mu_1 + \mu_2-a_1-1)\bigg]f =0.
\end{align}
Employing the M\"obius transformation,
$z=({\alpha\, r+\beta})/({\gamma\, r+\delta}),$ for $\alpha\delta-\beta\gamma\neq 0,$ yields
\begin{align} \label{sec2eq6}
-&\frac{(\alpha-\gamma z)^2 \left(\beta+\xi _1 \alpha-(\delta+\xi_1 \gamma) z\right) \left(\beta+\xi _2 \alpha-(\delta+\xi_2\gamma) z\right)}{(\alpha\delta-\beta \gamma)^2}\dfrac{d^2f(z)}{dz^2}\notag\\
&\hskip0.25true in+\bigg[\frac{(\alpha-\gamma z)[(b_1+2\, \mu _1\, \xi _2) (\alpha-\gamma z)+(2 \mu _1-a_1) (\beta-\delta z)]}{\alpha \delta-\beta \gamma}\notag\\
&\hskip0.5true in -\dfrac{2 (\alpha-\gamma z) \left(\beta-\delta z+(\alpha-\gamma z) \xi _1\right)((
\beta \gamma-\alpha \delta) \mu _2+\gamma \left(\delta z-\beta+(\gamma z-\alpha) \xi _2\right))}{(\alpha \delta-\beta \gamma)^2}\bigg]\dfrac{df(z)}{dz}\notag\\
&\hskip0.75true in+[\lambda -a_3- (\mu_1 + \mu_2)\, (\mu_1 + \mu_2-a_1-1)]\,f(z) =0.
\end{align}
Thus, if $\gamma = 0$, the change of variables $r\to z\equiv z(r)$ transforms \eqref{sec2eq6} into an equation of the same type as that of  \eqref{sec2eq5}. This in turn implies
\begin{align} \label{sec2eq7}
\left(z-\dfrac{\beta+\alpha \xi _1}{\delta} \right) \left(\dfrac{\beta+\alpha \xi _2}{\delta}- z\right)\dfrac{d^2f(z)}{dz^2}&+\left(z \left(a_1-2 \mu _1-2\mu _2\right)+\frac{\alpha b_1-\beta a_1+2 \mu _2 \left(\beta+\alpha \xi _1\right)+2\mu _1 \left(\beta+\alpha \xi _2\right)}{\delta}\right)\dfrac{df(z)}{dz}\notag\\
&+[\lambda -a_3- (\mu_1 + \mu_2)\, (\mu_1 + \mu_2-a_1-1)]\,f(z)=0.
\end{align}
To express the solutions of this equation in terms of hypergeometric functions, one must impose either of the following necessary conditions on $\alpha,\beta$ and $\delta$: 
$$(i)\hskip0.3true in \dfrac{\beta+\alpha \xi _1}{\delta} =0,\qquad \dfrac{\beta+\alpha \xi _2}{\delta}=1\quad\Longrightarrow\quad \alpha= \frac{\delta}{\xi _2-\xi _1},\quad \beta= \frac{\delta\, \xi _1}{\xi _1-\xi _2},$$
$$(ii)\hskip0.3true in \dfrac{\beta+\alpha \xi _1}{\delta} =1,\qquad \dfrac{\beta+\alpha \xi _2}{\delta}=0\quad\Longrightarrow\quad
\alpha= \frac{\delta}{\xi _1-\xi _2},\quad \beta= \frac{\delta\, \xi _2}{\xi_2-\xi _1}.$$
Should we impose the conditions as given $(i)$, equation \eqref{sec2eq7} reduces to (denoting $f \rightarrow \hat{f}$)
\begin{align}\label{sec2eq8}
z \left(1- z\right)\dfrac{d^2\hat{f}(z)}{dz^2}&+\left( \left(a_1-2 \mu _1-2\mu _2\right)\,z
+2\mu_1+\frac{b_1+ \xi _1 a_1}{\xi _2-\xi _1 }\right)\dfrac{d\hat{f}(z)}{dz}\notag\\
&+[\lambda -a_3- (\mu_1 + \mu_2)\, (\mu_1 + \mu_2-a_1-1)]\,\hat{f}(z)=0,
\end{align}
with two linearly independent series solutions  expressed in terms of the hypergeometric functions as
\begin{align} \label{sec2eq9}
&\hat{f}_1(z)= \pFq{2}{1} \left( \begin{array}{cc}
\hspace*{-0.2cm} \mu _1+ \mu
   _2-\frac{a_1+1}{2}-\sqrt{\left(\frac{a_1+1}{2}\right)^2-
   a_3+\lambda } \; , \quad  \mu _1+  \mu
   _2-\frac{a_1+1}{2}+\sqrt{\left(\frac{a_1+1}{2}\right)^2-
   a_3+\lambda }  \\ 
2\mu_1 +\frac{a_1 \xi_1+b_1 }{\xi_2-\xi_1} 
\end{array} \bigg| \; z\; \right),
\end{align}
and
\begin{align} \label{sec2eq10}
&\hat{f}_2(z)= z^{1-2 \mu _1+({b_1+a_1 \xi _1})/({\xi _1-\xi _2})}\notag\\
&\times\pFq{2}{1} \left( \begin{array}{cc}
\frac{1-a_1}{2}+\mu _2-\mu _1+\frac{b_1+a_1 \xi _1}{\xi _1-\xi _2}-\sqrt{\lambda - a_3+\left(\frac{1+a_1}{2}\right)^2}
 \; , \quad \frac{1-a_1}{2}+\mu _2-\mu _1+\frac{b_1+a_1 \xi _1}{\xi _1-\xi _2}+\sqrt{\lambda - a_3+\left(\frac{1+a_1}{2}\right)^2} \\ 
2-2 \mu _1-\frac{b_1+a_1 \xi_1}{\xi_2-\xi_1}
\end{array} \bigg| \; z\; \right).
\end{align}
Meanwhile, imposing the condition given in $(ii)$, equation \eqref{sec2eq7} reads
\begin{align}\label{sec2eq11}
z\left(1-z \right)\dfrac{d^2\breve{f}(z)}{dz^2}&+\left(\left(a_1-2 \mu _1-2\mu _2\right)\,z
+2\mu_2-\frac{b_1+  \xi _2 a_1}{\xi_2-\xi _1}\right)\dfrac{d\breve{f}(z)}{dz}\notag\\
&+[\lambda -a_3- (\mu_1 + \mu_2)\, (\mu_1 + \mu_2-a_1-1)]\,\breve{f}(z)=0.
\end{align}
with the two linearly independent solutions 
\begin{align} \label{sec2eq12}
\breve{f}_1(z)= \pFq{2}{1} \left( \begin{array}{cc}
\hspace*{-0.2cm} \mu_1+\mu_2-\frac{a_1+1}{2}+\sqrt{\left(\frac{a_1+1}{2}\right)^2+\lambda -a_3} \; , \; \mu_1+\mu_2-\frac{a_1+1}{2}-\sqrt{\left(\frac{a_1+1}{2}\right)^2+\lambda -a_3}  \\  
2\mu_2-\frac{a_1 \xi_2+b_1 }{\xi_2-\xi_1} 
\end{array} \Bigg| \; z\; \right),
\end{align}
and
\begin{align} \label{sec2eq13}
&\breve{f}_2(z)= z^{1-2 \mu _2+{(b_1+a_1 \xi _2)}/({\xi _2-\xi _1})}  \notag\\
&\times \pFq{2}{1} \left( \begin{array}{cc}
\frac{a_1+1}{2}+ \mu _1-\mu _2-\sqrt{\left(\frac{a_1+1}{2}\right)^2+\lambda -a_3}+\frac{b_1+a_1 \xi _1}{\xi _2-\xi _1}
 \; , \; \dfrac{a_1+1}{2}+ \mu _1- \mu _2+\sqrt{\left(\frac{a_1+1}{2}\right)^2+\lambda -a_3}+\frac{ b_1+a_1 \xi _1}{\xi _2-\xi _1}  \\  
2-2 \mu _2+\frac{b_1+a_1 \xi _2}{\xi _2-\xi _1}
\end{array} \Bigg| \; z\; \right).
\end{align}
More precisely, the analytic solutions of the differential equation \eqref{sec1eq4}, for arbitrary constants $A_1$ and $A_2$,
are
\begin{equation} \label{sec2eq14}
F(r)=\left( r - \xi_1 \right)^{\mu_1}\left( \xi_2-r \right)^{\mu_2} (A_1\hat{f}_1\left( r \right)+A_2 \hat{f}_2(r)),
\end{equation}
where
\begin{align} \label{sec2eq15}
&\hat{f}_1(r)= \pFq{2}{1} \left( \begin{array}{cc}
\hspace*{-0.2cm} \mu _1+ \mu
   _2-\frac{a_1+1}{2}-\sqrt{\lambda-a_3+\left(\frac{a_1+1}{2}\right)^2} \; , \;  \mu _1+  \mu
   _2-\frac{a_1+1}{2}+\sqrt{\lambda-a_3 + \left(\frac{a_1+1}{2}\right)^2}  \\ 
2\mu_1 +\frac{a_1 \xi_1+b_1 }{\xi_2-\xi_1} 
\end{array} \bigg| \; \frac{r-\xi _1}{\xi _2-\xi _1}\; \right),
\end{align}
and
\begin{align} \label{sec2eq16}
&\hat{f}_2(r)=z^{1-2 \mu _1+({b_1+a_1 \xi _1})/({\xi _1-\xi _2})}
\notag\\
&\times \pFq{2}{1} \left( \begin{array}{cc} \mu _2-\mu _1+\frac{1-a_1}{2}+\frac{b_1+a_1 \xi _2}{\xi _1-\xi _2}-\sqrt{\lambda - a_3+\left(\frac{1+a_1}{2}\right)^2} \; , \;  \mu _2- \mu _1+\frac{1-a_1}{2}+ \frac{b_1+a_1 \xi _2}{\xi _1-\xi _2}+\sqrt{\lambda - a_3+\left(\frac{1+a_1}{2}\right)^2} \\ 
\,2-2 \mu _1-\frac{b_1+a_1 \xi_1}{\xi_2-\xi_1}
\end{array} \bigg| \; \frac{r-\xi _1}{\xi _2-\xi _1}  \right).
\end{align}
Equivalently, the exact solutions of the differential equation \eqref{sec1eq4} may also be expressed as (denoting $f \rightarrow \breve{f}$)
\begin{equation} \label{sec2eq17}
F(r)=\left( r - \xi_1 \right)^{\mu_1}\left( \xi_2-r \right)^{\mu_2} (B_1\,\breve{f}_1\left( r \right)+B_2\, \breve{f}_2(r)),
\end{equation}
where
\begin{align} \label{sec2eq18}
\breve{f}_1(r)= \pFq{2}{1} \left( \begin{array}{c}
\hspace*{-0.2cm} \mu_1+\mu_2-\frac{a_1+1}{2}+\sqrt{\left(\frac{a_1+1}{2}\right)^2+\lambda -a_3} \; ,\quad \; \mu_1+\mu_2-\frac{a_1+1}{2}-\sqrt{\left(\frac{a_1+1}{2}\right)^2+\lambda -a_3}  \\ 
2\mu_2-\frac{a_1 \xi_2+b_1 }{\xi_2-\xi_1}  
\end{array} \Bigg| \; \frac{\xi _2-r}{\xi _2-\xi _1}\; \right),
\end{align}
and
\begin{align} \label{sec2eq19}
&\breve{f}_2(r)=z^{1-2 \mu _1+({b_1+a_1 \xi _1})/({\xi _1-\xi _2})}
\notag\\ 
&\times \pFq{2}{1} \left( \begin{array}{c}
 \mu _1-\mu _2+\frac{a_1+1}{2}-\sqrt{\left(\frac{a_1+1}{2}\right)^2+\lambda -a_3}+\frac{b_1+a_1 \xi _1}{\xi _2-\xi _1}\,,\, \quad \mu _1- \mu _2+ \frac{a_1+1}{2}+\sqrt{\left(\frac{a_1+1}{2}\right)^2+\lambda -a_3}+\frac{ b_1+a_1 \xi _1}{\xi _2-\xi _1}\\
2-2 \mu _2+\frac{b_1+a_1 \xi _2}{\xi _2-\xi _1}
\end{array} \Bigg| \; \dfrac{\xi _2-r}{\xi _2-\xi _1}\; \right) \, .
\end{align}
As a result, we are able to obtain \eqref{sec2eq14} from \eqref{sec2eq17} and vice versa. This relationship may be confirmed using the linear transformation \cite[Eq.15.8.4]{DLMF}):
\begin{align} \label{sec2eq20}
\dfrac{\sin(\pi(c-a-b))}{\pi}{}_2F_1\left(\begin{array}{c} a,b\\
c
\end{array}\big|z\right)&=\dfrac{1}{\Gamma(c-a)\Gamma(c-b)}\,{}_2F_1\left(\begin{array}{c}a,b\\ a+b-c+1\end{array} \bigg|1-z\right)\notag\\
&-\dfrac{(1-z)^{c-a-b}}{\Gamma(a)\Gamma(b)}\,{}_2F_1\left(\begin{array}{c} c-a,c-b\\
c-a-b+1\end{array}\big|1-z\right).
\end{align} 
After some simplification, it follows that
\begin{align}\label{sec2eq21}
&\frac{ \sin \left(\pi  \left(1-2 \mu _2-\frac{b_1+a_1 \xi _2}{\xi _1-\xi _2}\right)\right)}{\pi}\,\hat{f}_1(r)\notag\\
&=
\frac{\breve{f}_1(r)}{\Gamma \left(\mu _1-\mu _2+\frac{a_1+1}{2}-\sqrt{\left(\frac{1+a_1}{2}\right)^2+\lambda -a_3}+\frac{b_1+a_1 \xi _1}{\xi _2-\xi _1}\right) \Gamma \left(\mu _1-\mu _2+\frac{a_1+1}{2}+\sqrt{\left(\frac{1+a_1}{2}\right)^2+\lambda -a_3}+\frac{b_1+a_1 \xi _1}{\xi _2-\xi _1}\right)}\notag\\
&+\frac{(\xi_2-\xi_1)^{2 \mu _2-1-{(b_1+a_1 \xi _2)}/({\xi _2-\xi _1})}\,\breve{f}_2(r)}{ \Gamma \left(\mu _1+\mu _2-\frac{a_1+1}{2}- \sqrt{ \left(\frac{1+a_1}{2}\right)^2 +\lambda-a_3}\right) \Gamma \left(\mu _1+\mu _2-\frac{a_1+1}{2}+\sqrt{ \left(\frac{1+a_1}{2}\right)^2 +\lambda-a_3}\right)}\,.
\end{align}
For the second solution \eqref{sec2eq13}, we may again apply \eqref{sec2eq20} to obtain
\begin{align}\label{sec2eq22}
&\frac{\sin\left(\pi \left(1-2 \mu _2-\frac{b_1+ a_1 \xi _2}{\xi_1-\xi _2}\right)\right)}{\pi\, ((r-\xi_1)/(\xi_2-\xi_1))^{1-2 \mu _1+({b_1+a_1 \xi _1})/({\xi _1-\xi _2})}}\hat{f}_2(r)\notag\\
&=\dfrac{\, _2F_1\left(\begin{array}{c}
\frac{1-a_1}{2}- \mu _1+ \mu _2-\sqrt{(\frac{1+a_1}{2})^2+\lambda-a_3}+\frac{ b_1+a_1 \xi _1}{\xi _1-\xi _2},
\frac{1-a_1}{2}-\mu _1+\mu _2+\sqrt{(\frac{1+a_1}{2})^2+\lambda- a_3}+\frac{b_1+a_1 \xi _1}{\xi _1-\xi _2}\\
2\mu_2+\frac{b_1+a_1 \xi _2}{\xi _1-\xi _2}\\
\end{array}\bigg|
\frac{r-\xi _2}{\xi _1-\xi _2}\right)}
{\Gamma\left(\frac{3+a_1}{2}-\mu _1-\mu _2-\sqrt{(\frac{1+a_1}{2})^2+\lambda-a_3}\right) \Gamma \left(\frac{3+a_1}{2}-\mu _1-\mu _2-\sqrt{(\frac{1+a_1}{2})^2+\lambda-a_3}\right) }\notag\\
&-\dfrac{\, _2F_1\left(\begin{array}{c}
\frac{a_1+3}{2}-\sqrt{(\frac{1+a_1}{2})^2+\lambda-a_3}-\mu _1-\mu _2,\frac{a_1+3}{2}+\sqrt{(\frac{1+a_1}{2})^2+\lambda-a_3}-\mu _1-\mu _2\\
2-2 \mu _2-\frac{b_1+a_1 \xi _2}{\xi _1-\xi _2}\\
\end{array}\bigg|\frac{r-\xi _2}{\xi _1-\xi _2}\right)}
{\Gamma \left(\frac{1-a_1}{2}-2 \mu _1+2 \mu _2-\sqrt{\left(\frac{1+a_1}{2}\right)^2+\lambda-a_3}+\frac{b_1+a_1 \xi _1}{\xi _1-\xi _2}\right) \Gamma  \left(\frac{1-a_1}{2}-2 \mu _1+2 \mu _2+\sqrt{\left(\frac{1+a_1}{2}\right)^2+\lambda-a_3}+\frac{b_1+a_1 \xi _1}{\xi _1-\xi _2}\right) }\notag\\
&\times \left(\xi _2-r\right)^{1-2 \mu _2-\frac{b_1+a_1 \xi _2}{\xi _1-\xi _2}} \left({\xi _2-\xi _1}\right)^{1+2 \mu _2+\frac{b_1+a_1 \xi _2}{\xi _1-\xi _2}}.
\end{align} 
To obtain the expression in terms of $\breve{f}_1$ and $\breve{f}_2$, the Pfaff transformation \cite[Eq.15.8.1]{DLMF},
\begin{align} \label{sec2eq23}
{}_2F_1(a,b;c;z)=(1-z)^{c-a-b} {}_2F_1(c-a,c-b;c;z),
\end{align} 
must be used for the hypergeometric functions on the right-hand side of equation \eqref{sec2eq22}. It is sufficient to focus on either one of the solution sets. 
%%%%%%%%%%%%%%%%%%%%%%%%%%%%%%%%%%%%%%
\section{Connection with the Generalized Associated Legendre equation}
%%%%%%%%%%%%%%%%%%%%%%%%%%%%%%%%%%%
\noindent This section serves to demonstrate the relationship between the differential equation \eqref{sec1eq4} and equation \eqref{sec1eq3}. First denote
\begin{align}\label{sec3eq1}
k&= -\frac{1}{2}-\sqrt{\left(\frac{1+a_1}{2}\right)^2+\lambda - a_3},\quad n= 2 \mu _1-1+\frac{b_1+a_1 \xi _1}{\xi_2-\xi _1},\quad m= 1-2 \mu _2-\frac{b_1+a_1 \xi _2}{\xi _1-\xi _2},
\end{align}
then, using partial-fraction decomposition, equation \eqref{sec1eq4} reads
\begin{align}\label{sec3eq2}
\left( r-\xi_1 \right) \left(\xi_2-r \right)&\frac{d^2F(r)}{dr^2}+ \big( r \left(m-n-2+2 \mu _1+2 \mu _2\right)+\left(1-m-2 \mu _2\right) \xi _1+\left(1+n-2 \mu _1\right) \xi _2\big) \frac{dF(r)}{dr} \notag\\
&+ \left( \left(k+1+\frac{n-m}{2}-\mu _1-\mu _2\right) \left(k+\frac{m-n}{2}+\mu _1+\mu _2\right)+\frac{b_2+c_3+a_2 \xi _1+b_3 \xi _1+a_3 \xi _1^2}{\left(r-\xi _1\right) \left(\xi _2-\xi _1\right)}\right.\notag\\
&\left.+\frac{b_2+c_3+a_2 \xi _2+b_3 \xi _2+a_3 \xi _2^2}{\left(r-\xi _2\right) \left(\xi _1-\xi _2\right)} \right)F(r) = 0.
\end{align}
Using \eqref{sec3eq1}, the indicial equations \eqref{sec2eq1} and \eqref{sec2eq2} may be expressed as 
\begin{align*}
a_3 \xi _1^2+(a_2+b_3) \xi _1+b_2+c_3&=\mu _1 \left(\mu _1-n\right) \left(\xi _1-\xi _2\right)^2 ,
\quad a_3 \xi _2^2+(a_2+b_3) \xi _2+b_2+c_3=\mu _2 \left(m+\mu _2\right) \left(\xi _1-\xi _2\right)^2 \, .\end{align*}
 Thus, equation \eqref{sec3eq2} reduces to
\begin{align}\label{sec3eq3}
\left( r-\xi_1 \right) \left(\xi_2-r \right)\frac{d^2F(r)}{dr^2} &+ \big( r \left(m-n-2+2 \mu _1+2 \mu _2\right)+\left(1-m-2 \mu _2\right) \xi _1+\left(1+n-2 \mu _1\right) \xi _2\big) \frac{dF(r)}{dr} \notag\\
&+ \left( \left(1+k+\frac{n-m}{2}-\mu _1-\mu _2\right) \left(k+\frac{m-n}{2}+\mu _1+\mu _2\right)\right.\notag\\
&\hskip1true in \left.+\frac{\mu _1 \left(\mu _1-n\right) \left(\xi _1-\xi _2\right)}{\left(r-\xi _1\right) }+\frac{\mu _2 \left(m+\mu _2\right) \left(\xi _1-\xi _2\right)}{\left(r-\xi _2\right)} \right)F(r) = 0,
\end{align}
with the exact solutions, determined via \eqref{sec2eq17}, given by: 
\begin{align}
F_1(r)&=\left( r - \xi_1 \right)^{\mu_1}\left( \xi_2-r \right)^{\mu_2} \,{}_2F_1\left(
\begin{array}{c}
-k+\tfrac{n-m}{2},\quad k+1+\tfrac{n-m}{2}\\
1-m
\end{array}\bigg|\frac{\xi_2-r}{\xi _2-\xi _1}\right),\label{sec3eq4}\\
F_2(r)&=\left( r - \xi_1 \right)^{\mu_1}\left( \xi_2-r \right)^{\mu_2+m} \, _2F_1\left(
\begin{array}{c}
-k+\frac{n+m}{2},\quad k+1+\frac{n+m}{2}\\
1+m
\end{array}\bigg|\frac{\xi_2-r}{\xi _2-\xi _1}\right),\label{sec3eq5}
\end{align}
as found earlier by Kuipers \emph{et al.} for $\mu_1=n/2$ and $\mu_2=-m/2$. In this case equation \eqref{sec3eq3} reads
\begin{align}\label{sec3eq6}
&\left( r-\xi_1 \right) \left(\xi_2-r \right)\frac{d^2F(r)}{dr^2} + \left(-2 r+\xi _1+\xi _2\right) \frac{dF(r)}{dr}+ \left(k (k+1)+\frac{n^2 \left(\xi _1-\xi _2\right)}{4 \left(r-\xi _1\right)}-\frac{m^2 \left(\xi _1-\xi _2\right)}{4 \left(r-\xi _2\right)}\right)F(r) = 0.
\end{align}
For $\xi_2=-\xi_1=1$, equation \eqref{sec3eq6} reduces to the Generalized Associated Legendre Differential Equation. Different choices of $\xi_1$ and $\xi_2$ give rise to other interesting classes of differential equations, (e.g. $ \, \xi_1=0 \, , \, \xi_2=1$). The mathematical properties of these other classes (such as the weight-function, the recurrence relation, the orthogonality conditions, etc.) will be the focus of future work. 
%%%%%%%%%%%%%%%%%%%%%%%%%%%%%%%%%%%%%%
\section{Connection with the Universal Associated Legendre equation}
%%%%%%%%%%%%%%%%%%%%%%%%%%%%%%%%%%%
\noindent To establish the connection between the solutions of the differential equation \eqref{sec1eq4} and that of the Universal Associated Legendre Equation \eqref{sec1eq2} as given by \eqref{sec1eq1}, we express \eqref{sec1eq1} in terms of the hypergeometric function. Using the Legendre Duplication Formula, 
$\Gamma(2z)=2^{2z-1}\Gamma(z)\Gamma(z+\tfrac12)/\sqrt{\pi}$,
equation \eqref{sec1eq1} can be written as
\begin{align}\label{sec4eq1}
F(r)&=\sqrt{\frac{(2\,\ell'+1)\,(\ell'-m')!}{2\,\Gamma(\ell'+m'+1)}}(1-r^2)^{\frac{m'}2}\sum_{\nu=0}^{[\frac12(\ell'-m')]}\dfrac{(-1)^\nu 2^{m'} \Gamma(\ell'-\nu+\tfrac12)}{ \nu!\,\Gamma(\frac{\ell'-m'+1}{2}-\nu)\,\Gamma(\frac{\ell'-m'}{2}-\nu+1)}\,r^{\ell'-m'-2\nu} \, .
\end{align}
Further, by means of the Pochhammer identity $\Gamma(z-\nu)={(-1)^\nu\Gamma(z)}/{(1-z)_\nu}$, we obtain
\begin{align}\label{sec4eq2} 
F(r)
&=\dfrac{2^{m'} \Gamma(\ell'+\tfrac12)}{\Gamma(\frac{\ell'-m'+1}{2})\Gamma(\frac{\ell'-m'}{2}+1)}\sqrt{\frac{(2\ell'+1)(\ell'-m')!}{2\Gamma(\ell'+m'+1)}}r^{\ell'-m'}\,(1-r^2)^{\frac{m'}2}\sum_{\nu=0}^{[\frac12(\ell'-m')]}\dfrac{(\frac{1-(\ell'-m')}{2})_{\nu} (-\frac{\ell'-m'}{2})_{\nu}}{ \nu!(\frac12-\ell')_{\nu}}r^{-2\nu}.
\end{align}
With the assumption that $-\frac12(\ell'-m')$ or $\frac12({1-(\ell'-m')}) =0,-1,-2,\dots,$ this equation may now be written in terms of the hypergeometric equation 
\begin{align}
F(r)
&=\dfrac{2^{m'-1/2}\Gamma(\ell'+\frac12)}{\Gamma(\frac{\ell'-m'+1}{2})\Gamma(\frac{\ell'-m'}{2}+1)}\sqrt{\frac{(2\ell'+1)(\ell'-m')!}{\Gamma(\ell'+m'+1)}}r^{\ell'-m'}\,(1-r^2)^{\frac{m'}{2}}{}_2F_1\left(\begin{array}{c}
\frac{1-(\ell'-m')}{2},-\frac{\ell'-m'}{2}\\
\frac12-\ell'\end{array}\bigg|\frac{1}{r^{2}}\right),
\end{align}
with the understanding that the limit of the right-hand side is well-defined as $r \rightarrow 0$. Since $$\Gamma\left(\frac{\ell'-m'+1}{2}\right)\Gamma\left(\frac{\ell'-m'}{2}+1\right)=2^{m'-\ell'}\sqrt{\pi}\, \Gamma (\ell'-m'+1),$$ it easily follows that
\begin{align}
F(r)
&=\dfrac{2^{\ell'-\frac12}\Gamma(\ell'+\frac12)}{\sqrt{\pi} }\sqrt{\frac{(2\ell'+1)}{(\ell'-m')!\,\Gamma(\ell'+m'+1)}}r^{\ell'-m'}\,(1-r^2)^{\frac{m'}{2}}
{}_2F_1\left(\begin{array}{c}
\frac{1-(\ell'-m')}{2},-\frac{\ell'-m'}{2} \\
\frac12-\ell'\end{array}\bigg|\frac{1}{r^{2}}\right).
\end{align}
The identity \cite[Eq.15.8.6]{DLMF}
$$
(-1)^m\dfrac{(c)_m}{(b)_m}\,{}_2F_1\left(\begin{array}{c}-m,b\\
c\end{array}\bigg|z\right)=z^m{}_2F_1\left(\begin{array}{c}-m,1-c-m\\
1-b-m\end{array}\bigg|\dfrac1z\right),\qquad m=0,1,2,\dots,
$$
implies
\begin{align} \label{sec4eq5}
F(r)
&=\dfrac{(-1)^{\frac{\ell'-m'}{2}} 2^{\ell'-\frac12}{\Gamma(\ell'+\frac12)} \left(\frac{1}{2}\right)_{\frac{\ell'-m'}{2}}}{\sqrt{\pi}\left(\frac{1+\ell'+m'}{2}\right)_{\frac{\ell'-m'}{2}}}
 \sqrt{\tfrac{(2\ell'+1)}{(\ell'-m')!\Gamma(\ell'+m'+1)}} (1-r^2)^{\frac{m'}2} {}_2F_1\left(\frac{1+\ell'+m'}{2},-\frac{\ell'-m'}{2} ;\frac12;r^{2}\right).
\end{align}
With $a_1=2(\mu_1+\mu_2-1)$, the differential equation \eqref{sec1eq4} reads after an application of partial-fraction decomposition,
 \begin{align}\label{sec4eq6}
\left(\xi_2-r \right)\left( r-\xi_1 \right) \frac{d^2F(r)}{dr^2}& + \left( 2 \left(\mu _1+\mu _2-1\right) r + b_1 \right) \frac{dF(r)}{dr}\notag\\
& + \left( \lambda -a_3 + \frac{a_3 \xi _1^2+(a_2+b_3) \xi _1+b_2+c_3}{\left(\xi _2-\xi _1\right) \left(r-\xi _1\right)}+ \frac{a_3 \xi _2^2+(a_2+b_3) \xi _2+b_2+c_3}{\left(\xi _2-\xi _1\right) \left(\xi _2-r\right)} \right)F(r) = 0.
\end{align}
Using the identity \cite[Eq.15.8.20]{DLMF}:
$$
{}_2F_1\left(
\begin{array}{c} a, 1-a \\ 
c\end{array} \bigg|z\right)=(1-z)^{c-1} {}_2F_1\left(\begin{array}{c}\tfrac{c-a}{2},
\tfrac{a+c-1}{2}\\
c\end{array}\bigg|4z(1-z)\right)
$$
the solution of \eqref{sec4eq6}, namely,
\begin{align} \label{sec4eq7}
F(r)&=\left( r - \xi_1 \right)^{\mu_1}\left( \xi_2-r \right)^{\mu_2} \notag\\
&\times \pFq{2}{1} \left( \begin{array}{cc}
\hspace*{-0.2cm} \mu _1+ \mu
   _2-\frac{a_1+1}{2}-\sqrt{\left(\frac{a_1+1}{2}\right)^2-
   a_3+\lambda } \; , \;  \mu _1+  \mu
   _2-\frac{a_1+1}{2}+\sqrt{\left(\frac{a_1+1}{2}\right)^2-
   a_3+\lambda }  \\ 
2\mu_1 +\frac{a_1 \xi_1+b_1 }{\xi_2-\xi_1} 
\end{array} \bigg| \; \frac{r-\xi _1}{\xi _2-\xi _1}\; \right).
\end{align}
can be written as
\begin{align}\label{sec4eq8}
&F(r)=\left(r-\xi _1\right)^{\mu _1} \left(\xi _2-r\right)^{\mu _2} \left(\frac{r-\xi _1}{\xi _2-\xi _1}\right)^{1-\frac{b_1+2 \mu _2 \xi _1+2 \left(\mu _1-1\right) \xi _2}{\xi _1-\xi _2}}\notag\\
& _2F_1\left(\begin{array}{c}
\frac{2 b_1+\left(4 \mu _2-1\right) \xi _1+\left(4 \mu _1-3\right) \xi _2}{4 \left(\xi _1-\xi _2\right)}+\frac{\sqrt{\lambda -a_3+\left(\mu _1+\mu _2-\frac12\right)^2}}{2},\frac{2 b_1+\left(4 \mu _2-1\right) \xi _1+\left(4 \mu _1-3\right) \xi _2}{4 \left(\xi _1-\xi _2\right)}
-\frac{\sqrt{\lambda -a_3+ \left( \mu _1+ \mu _2-\frac12\right)^2}}{2}\\
\frac{b_1+2 \mu _2 \xi _1+2 \left(\mu _1-1\right) \xi _2}{\xi _1-\xi _2}
\end{array}\bigg|\frac{4 \left(r-\xi _1\right) \left(\xi _2-r\right)}{\left(\xi _1-\xi _2\right)^2}\right) \, .
\end{align}
Using the identity   \cite[Eq.15.8.7]{DLMF},
$$
{}_2F_1\left(\begin{array}{c} -m,\quad\gamma\\
\gamma-c-m+1\end{array}\bigg| 1-z\right)=\frac{(c)_m}{(c-\gamma)_m}{}_2F_1\left(\begin{array}{c}-m,\gamma\\
c\end{array}\bigg|z\right)$$
and assuming 
\begin{align*}
\frac{2 b_1+\left(4 \mu _2-1\right) \xi _1+\left(4 \mu _1-3\right) \xi _2}{4 \left(\xi _1-\xi _2\right)}+\frac{\sqrt{\lambda -a_3+\left(\mu _1+\mu _2-\frac12\right)^2}}{2}&=-\frac{\ell'-m'}{2},\\
\frac{2 b_1+\left(4 \mu _2-1\right) \xi _1+\left(4 \mu _1-3\right) \xi _2}{4 \left(\xi _1-\xi _2\right)}
-\frac{\sqrt{\lambda -a_3+ \left( \mu _1+ \mu _2-\frac12\right)^2}}{2}&=\dfrac{1+\ell'+m'}{2},
\end{align*}
it follows that $c=1/2$. Finally, the solution \eqref{sec4eq8} now reads
\begin{align}\label{sec4eq9} 
F(r)&=\left(\frac{r-\xi _1}{\xi _2-\xi _1}\right)^{1-\frac{b_1+2 \mu _2 \xi _1+2 \left(\mu _1-1\right) \xi _2}{\xi _1-\xi _2}}
\dfrac{\left(\frac12\right)_{\frac{\ell'-m'}{2}}\left(r-\xi _1\right)^{\mu _1} \left(\xi _2-r\right)^{\mu _2} }{\left(-\dfrac{\ell'+m'}{2}\right)_{\frac{\ell'-m'}{2}}}  \,_2F_1\left(\begin{array}{c}
-\tfrac{\ell'-m'}{2},\tfrac{1+\ell'+m'}{2}\\
\tfrac12\end{array}\bigg|\frac{\left(\xi _1+\xi _2-2r\right)^2}{\left(\xi _1-\xi _2\right){}^2}\right)
\end{align}
which reduces, up to a multiplicative constant, to the solution \eqref{sec4eq5} for $b_1=0,~\xi_1=-1$ and $\xi_2=1$.
%-----------------------
\section{Conclusion}
% ------------------------------------------------------
\vs The classical Generalized and the recent Universal Associated Legendre Equations are members of the more broad class of differential equations given by \eqref{sec1eq4}. We established the hypergeometric solutions of this class of equations and demonstrated that they lead to the Generalized and Universal Associated Legendre hypergeometric solutions. These new solutions open the door for further compelling studies, including the examination of their mathematical properties and the investigation of their applicability to problems in mathematical physics.
%------------------------------------------------------
\section{Acknowledgments}
% ------------------------------------------------------
\medskip
\noindent Partial financial support of this work under Grant No. GP249507 from the
Natural Sciences and Engineering Research Council of Canada is gratefully acknowledged.
% --------------------------------------------------------------------------------

\end{document}